\documentclass[aps,prl,superscriptaddress,twocolumn,showpacs]{revtex4}
\usepackage{graphics}
\usepackage{epsfig}
\usepackage{graphicx}
\usepackage{amsthm} % for theorems
\usepackage{amsbsy} % for bold symbols (\bf doesn't work for Greek letters)
\usepackage{dcolumn}   % Align table columns on decimal point
\usepackage{verbatim}   % useful for program listings
\usepackage{color}     	 % use if color is used in text
\usepackage{subfigure}  % use for side-by-side figures
\usepackage{amsmath,amsfonts,amssymb,graphics,graphicx,epsfig,color,times,natbib}

\theoremstyle{remark}

\theoremstyle{definition}

\newcommand{\ie}{{\it i.e.\,}}

\newcommand{\bra}[1]{\left\langle#1\right|} % } These resize
\newcommand{\ket}[1]{\left|#1\right\rangle} % }
\newcommand{\icfo}{\affiliation{ICFO--Institut de Ci\`encies Fot\`oniques, E-08860 Castelldefels, Barcelona, Spain}}
\newcommand{\icrea}{\affiliation{ICREA--Instituci\'o Catalana de Recerca i Estudis Avan\c{c}ats, Lluis Companys 23, 08010 Barcelona,
Spain}}
\newcommand{\bristol}{\affiliation{H.H. Wills Physics Laboratory, University of Bristol, Bristol BS8 1TL, United Kingdom}}

\begin{document}
%Assessing the Hilbert space dimension of arbitrary quantum systems\\ or \\
\title{Device-independent tests of classical and quantum dimensions}
\author{Rodrigo Gallego}\icfo
\author{Nicolas Brunner}\bristol
\author{Christopher Hadley}\icfo
\author{Antonio Ac\'in}\icfo \icrea
%\pacs{75.10.Pq,	03.65.Ud, 03.67.-a}
\date{\today}

\begin{abstract}
We address the problem of testing the dimensionality of classical and quantum systems in a `black-box' scenario. We develop a general formalism for tackling this problem. This allows us to derive lower bounds on the classical dimension necessary to reproduce given measurement data. Furthermore, we generalise the concept of quantum dimension witnesses to arbitrary quantum systems, allowing one to place a lower bound on the Hilbert space dimension necessary to reproduce certain data. Illustrating these ideas, we provide simple examples of classical and quantum dimension witnesses.
\end{abstract}
\maketitle

%%%%%%%%%%%%%%%%%%%%%%%%%%%%%%%%%%%
% Introduction
%%%%%%%%%%%%%%%%%%%%%%%%%%%%%%%%%%

In quantum mechanics, experimental observations are usually described using theoretical models which make specific assumptions on the physical system under consideration, including the size of the associated Hilbert space.  The Hilbert space dimension is thus intrinsic to the model.  In this work, the converse approach is considered: is it possible to assess the Hilbert space dimension from experimental data without an {\it a priori} model?

This is particularly relevant in the context of quantum information science, in which dimensionality enjoys the status of a resource for information processing.  Higher dimensional systems may potentially enable the implementation of more efficient and powerful protocols.  It is therefore desirable to design methods for testing the Hilbert space dimension of quantum systems which are `device-independent'; that is, where no assumption is made on the devices used to perform the tests.

Recent years have seen the problem of testing the dimension of a non-characterised system considered from different perspectives. Initially, the concept of a dimension witness was introduced by Brunner {\it et al.} \cite{dimH} in the context of non-local correlations.  Such witnesses are essentially Bell-type inequalities, the violation of which imposes a lower bound on the Hilbert space dimension of the entangled state on which local measurements have been performed \cite{perezgarcia,vertesi1,vertesi2,vertesi3,briet,vertesi4,junge}.  Wehner {\it et al.} \cite{wehner} subsequently showed how the problem relates to random-access codes, and could thus exploit previously known bounds.  Finally, Wolf and Perez-Garcia \cite{wolf} addressed the question from a dynamical viewpoint, showing how bounds on the dimensionality may be obtained from the evolution of an expectation value.

Though these works represent significant progress, they all have substantive drawbacks. The approach of Ref. \cite{dimH} may not be applied to single-party systems as it is based on the non-local correlations between distant
particles; the bounds of Ref. \cite{wehner} are based on Shannon channel capacities, which are, in general, difficult to compute; whilst the approach of Ref. \cite{wolf} cannot be applied to the static case. More generally, all these works show how to
adapt existing techniques developed for other scenarios to the problem of assessing the dimension of a non-characterised system. However, (i) no systematic approach to this problem has yet been developed and (ii) there are no techniques specifically designed to tackle this question.

In the present work we bridge this gap and formalise the problem of testing the Hilbert space dimension of arbitrary quantum systems in the simplest scenarios in which the problem is meaningful.  We introduce natural tools for addressing the problem, starting by developing methods for determining the minimal dimensionality of classical systems, given certain data.  Using geometrical ideas, we introduce the idea of {\it tight classical dimension witnesses}, leading to a generalisation of quantum dimension witnesses to arbitrary systems.  As an illustration of our general formalism, we provide simple examples of such classical and quantum dimension witnesses.

\begin{small}\begin{figure}
 \begin{center}\begin{tabular}{c}
  \includegraphics[width=7cm,angle=0]{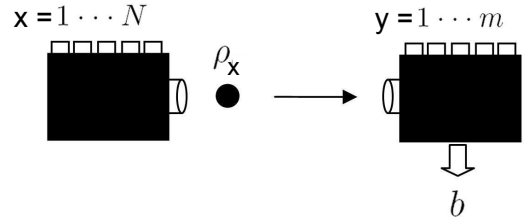}
 \end{tabular}\end{center}
 \caption{Device-independent test of classical or quantum dimensionality. Our scenario features two black boxes: a state preparator and a measurement device.}
 \label{strategydiag}
\end{figure}\end{small}

\textbf{\emph{Scenario.}}---We consider the scenario depicted in Fig. 1.  An initial `black box', the {\it state preparator}, prepares upon requests a state---we will consider the case of both classical and quantum states. The box features $N$ buttons which label the prepared state; when pressing button $x$, the box emits the state $\rho_x$ where $x \in \{1,...,N\}$. The prepared state is then sent to a second black box, the \textit{measurement device}.  This box performs a measurement $y \in \{1,...,m\}$ on the state, delivering outcome $b \in \{1,...,k\}$. The experiment is thus described by the probability distribution $P(b|x,y)$, giving the probability of obtaining outcome $b$ when measurement $y$ is performed on the prepared state $\rho_x$.

Our goal is to estimate the minimal dimension of the
mediating particle between the devices needed to describe the
observed statistics. That is, what are the minimal classical and
quantum dimensions necessary to reproduce a given set of
probabilities $P(b|x,y)$?

Formally, a probability distribution $P(b|x,y)$ admits a $d$-dimensional quantum representation if it can be written in the form
\begin{equation}\label{Q}
P(b|x,y)=\textrm{tr}(\rho_{x} M_{b}^{y}),
\end{equation}
for some state $\rho_x$ and operators $M_b^y$ acting on $\mathbb{C}^d$. We also say that $P(b|x,y)$ has a classical $d$-dimensional representation if it can be written
\begin{equation}\label{C}
P(b|x,y)=P(b|\Lambda_x ,y),
\end{equation}
where $\Lambda_x$ is a classical state of dimension $d$, {\it i.e.} a probability distribution $\vec{q}$ over classical dits, where $q_j=P(\Lambda_x=j)$ and $\sum_j q_j=1$.  The outcome $b$ is then a function of the state $\Lambda_x$ and the measurement $y$. This model is in the spirit of ontological models, recently investigated in Refs. \cite{harrigan,galvao}.

\textbf{\emph{Tight classical dimension witnesses.}}---We start by deriving a general method for finding a lower bound on the dimensionality of the classical states $\Lambda_x$ necessary to reproduce a given probability distribution $P(b|x,y)$.  For simplicity we shall focus on measurements with binary outcomes, which we denote $b=\pm 1$; the generalisation to larger alphabets is straightforward. It then becomes convenient to use expectation values:
\begin{equation}\label{mean}
E_{xy}= P(b=+1|x,y)-P(b=-1|x,y ).
\end{equation}
Every experiment is characterised by a vector of correlation functions
\begin{equation}
\vec{E} = ( \vec{v}_{x=1} , \vec{v}_{x=2}, ..., \vec{v}_{x=N} ),
\end{equation}
where $\vec{v}_{x}= (E_{x1},E_{x2},...,E_{xm})$ is a vector containing the correlation functions for a given preparation $x$ and all measurements. Deterministic experiments---those in which only one outcome appears for any possible pair of preparation and measurement---correspond to vectors $\vec{E}_{\rm det}$ for which $E_{xy}=\pm1$ for all $x,y$. Clearly, any possible experiment may be written as a convex combination of deterministic vectors $\vec{E}_{\rm det}$. Thus, the set of all possible experiments defines a polytope---\ie a convex set with a finite number of extremal points---denoted in what follows by $\mathbb{P}_{N,m}$. The facets of $\mathbb{P}_{N,m}$ are termed positivity facets, of the form $E_{xy}\leq 1$ and $E_{xy}\geq -1$, which ensures that probabilities $P(b|x,y)$ are well defined. Thus $\mathbb{P}_{N,m}$ may be viewed as the set of all valid probability distributions. Note that $\mathbb{P}_{N,m}$ resides in a space of dimension $Nm$ and has $2^{Nm}$ vertices, corresponding to the deterministic vectors $\vec{E}_{\rm det}$.

Next, we would like to characterise the set of realisable experiments in the case that the dimension $d$ of the classical states is limited.  We first note that if $d\geq N$, all possible experiments can be realised.  Indeed, it is then possible to encode the choice of preparation $x$ in the classical state $\Lambda_x$; {\it i.e.} $\Lambda_x=x$.  Thus, any probability distribution $P(b|x,y)$---{\it i.e.} any vector $\vec{E}$ in $\mathbb{P}_{N,m}$---can be obtained, since the measurement device has full information of both $x$ and $y$.

Therefore the problem of bounding the dimension of classical (or quantum) systems necessary to reproduce a given set of data is meaningful only if $d<N$.  In this case, it turns out that not all possible experiments can be realised.  Let us first focus on deterministic experiments.  Clearly, if the classical state sent by the state preparator is of dimension $d<N$, then (at least) $\lceil N/d \rceil$ preparations must correspond to the same state ({\it i.e.} the same classical dit). Therefore, only a subset of the $2^{Nm}$ deterministic vectors can be obtained in this case: those deterministic vectors $\vec{E}_{\rm det}^{d}$ composed of (at least) $\lceil N/d\rceil$ vectors $\vec{v}_x$ which are the same.

General strategies consist of mixtures of these deterministic points. It is however possible to identify two different scenarios. In the first scenario, the state preparator and the measurement device share no pre-established correlations and, thus, mix different deterministic preparations and measurements in an uncorrelated manner. In a practical setup, this is often a very reasonable assumption. In this case, the set of experiments is not convex, as not every mixture of points $\vec{E}_{\rm det}^{d}$ is realisable with systems of dimension $d$~\cite{next}. In the second scenario, the state preparator and the measurement device share classical correlations. This is the natural situation in a device-independent scenario, where no assumption about the devices is possible. Now, the set of realisable points is by construction convex and corresponds to the convex hull of deterministic vectors $\vec{E}_{\rm det}^{d}$, a polytope denoted $\mathbb{P}_{N,m}^d$. In this work, we focus on the second scenario since: (i) its characterisation is simpler, as a polytope is defined by a finite set of linear inequalities and (ii) it is more general, as any experiment in the first scenario is contained in $\mathbb{P}_{N,m}^d$.

The polytope $\mathbb{P}_{N,m}^d$ is a strict subset of $\mathbb{P}_{N,m}$. Thus it features additional facets which are not positivity facets. These new facets are `tight classical dimension witnesses' (for systems of dimension $d$), and are formally given by linear combinations of the expectation values $E_{xy}$; {\it i.e.} \begin{equation}\label{dimwit}
\vec{W} \cdot \vec{E} =\sum_{x,y}w_{xy} E_{xy} \le C_d
\end{equation}
where the probabilities (entering $E_{xy}$) are of the form of Eq. \eqref{C} with $\Lambda_x$ being a classical state of dimension $d$. These inequalities are classical dimension witnesses in the sense that: (i) for any experiment involving classical states of dimension $d$, the associated correlation vector $\vec{E}$ will satisfy inequality \eqref{dimwit}; (ii) in order to violate inequality \eqref{dimwit}, classical systems of dimension strictly larger than $d$ are required. Note that a witness is termed `tight' when it corresponds to a facet of the polytope $\mathbb{P}_{N,m}^{d}$; this terminology is borrowed from the study of non-locality, in analogy to tight Bell inequalities.

To summarise, by characterising the polytopes $\mathbb{P}_{N,m}^d$ (that is, by finding all the facets of $\mathbb{P}_{N,m}^d$) one can lower bound the dimension of classical system necessary to reproduce a given probability distribution $P(b|x,y)$. Clearly, if a probability distribution is proven not to belong to $\mathbb{P}_{N,m}^d$, it requires classical systems of dimension strictly larger than $d$. In the case that the state preparator and the measuring device are allowed to share pre-established correlations, our technique also provides an upper bound on the dimension, since all experiments in $\mathbb{P}_{N,m}^d$ can then be obtained from classical systems of dimension $d$. In this case our methods makes it possible, in principle, to determine the minimum dimensionality required in order to reproduce any given probability distribution.

\textbf{\emph{Quantum dimension witnesses.}}---The above ideas can be extended to the problem of finding lower bounds on the Hilbert space dimension of quantum systems necessary to reproduce a certain probability distribution. We first define linear quantum $d$-dimensional witnesses as linear expression of the form
\begin{equation}
\vec{W} \cdot \vec{E} =\sum_{x,y}w_{xy} E_{xy} \le Q_d,
\end{equation}
where the correlation functions $E_{xy}$ can be written in terms of probabilities of the form \eqref{Q} with $\rho_x$ acting on $\mathbb{C}^d$, and there exists a probability distribution $P(b|x,y)$ such that $\vec{W}\cdot \vec{E}>Q_d$. This generalises the concept of dimension witness of Ref. \cite{dimH} to arbitrary quantum systems.

It would be, in general, very interesting to fully characterise the set of experiments, {\it i.e.} of vectors $\vec{E}$, that can be obtained from quantum states of a given dimension. Indeed, this would allow one to determine the minimal Hilbert space dimension necessary to reproduce any given probability distribution. As above, it is possible to define different scenarios, depending on whether the state preparator and the measurement device share correlations, which can now be quantum. In the case of no correlations, the set of realisable points is again not convex~\cite{next}. In the case of correlated devices, the set of quantum experiments is convex. However, obtaining its complete characterisation represents a more difficult problem, since it is not a polytope. That is, the number of extreme points is infinite and its boundary cannot be characterised by a finite number of linear dimension witnesses. All these different scenarios will be discussed elsewhere \cite{next}. As stated, for the sake of simplicity, our analysis here is restricted to devices sharing classical correlations.

\textbf{\emph{Case studies.}}---As an application of our general formalism, we now present several examples of dimension witnesses. In particular, we give a family of linear witnesses which can be used as both a classical and quantum witness for any dimension. In general, the classical and quantum bounds of our witnesses---$C_d$ and $Q_d$, respectively---differ, and thus our witnesses can distinguish between classical and quantum resources of given dimensions. We also give an example of a non-linear witness for qubits. 

\emph{1. Simplest case.} We start by considering the case $d=2$, {\it i.e.} where the classical state sent by the state preparator is simply a bit. Indeed, we saw above that our problem is meaningful only if $d<N$, and thus we consider the case of three preparations ($N=3$) and two measurements ($m=2$) with binary outcomes \footnote{Note that the CHSH polynomial does not work as a dimension witness, since it features only two preparations---indeed, it is necessary to have $d<N$. In the device-independent scenario considered here, it is possible to reach the maximum of CHSH=4 by sending a classical bit (the bit simply indicates which preparation has been chosen).}. We fully characterise the polytope $\mathbb{P}_{3,2}^2$. It features a single type of non-trivial facet given by
\begin{equation}\label{322}
 I_3 \equiv | E_{11}+E_{12}+E_{21}-E_{22}-E_{31} | \leq 3.
\end{equation}
This inequality is a tight 2-dimensional classical witness. To be violated, trits (or higher-dimensional systems) are required. Note that trits are sufficient to reach the algebraic maximum of $I_3=5$; indeed any correlation vector $\vec{E}$ in $\mathbb{P}_{3,2}$ can be obtained using trits. Fig. 2 shows a schematic view of the situation.

\begin{small}\begin{figure}
 \begin{center}\begin{tabular}{c}
  \includegraphics[width=7cm,angle=0]{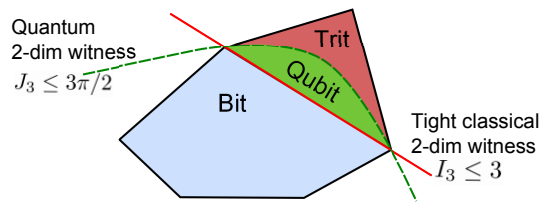}
 \end{tabular}\end{center}
 \caption{Schematic representation of the sets of experiments achievable from classical and quantum states of given dimensions for case study 1. The set of experiments (more precisely its convex hull) attainable from 2-dimensional classical states, {\it i.e.} bits, forms the polytope $\mathbb{P}_{3,2}^2$ (blue region). The inequality $I_3\leq3$ (solid line), a facet of this polytope, is a `tight 2-dimensional classical witness'. The set of experiments attainable from 2-dimensional quantum states, {\it i.e.} qubits, (green and blue region) is strictly larger. The inequality $J_3\leq \frac{3\pi}{2}$ (dashed curve) is a qubit-witness; it cannot be violated by performing measurements on qubits: qutrits are required. The set of all possible experiments (blue, green and red regions) forms the polytope $\mathbb{P}_{3,2}$; any point in it can be reproduced with a trit or a qutrit.}
 \label{strategydiag}
\end{figure}\end{small}

The witness $I_3$ is also a 2-dimensional quantum witness. The maximal value of $I_3$ obtainable from qubits can be computed analytically. Here the analysis may be restricted to pure states, since $I_3$ is a linear expression of the probabilities, and to rank-one projective measurements, since we consider measurements of two outcomes \cite{masanes}. By solving the maximisation problem, it can be shown that $\max_{\rho\in\mathcal{B}(\mathbb{C}^{2})}I_3 = 1+2\sqrt{2} \approx3.8284$. The first four terms in Eq. (\ref{322}) can be seen as the CHSH polynomial, whose maximum quantum value is equal to $2\sqrt{2}$. This maximisation does not involve the third preparation, which can be always chosen such that $E_{31}=-1$. In order to quantum mechanically reproduce a probability distribution $P(b|x,y)$ leading to $I_3>1+2\sqrt{2}$, qutrits (or systems of higher dimension) are required; in fact classical trits would suffice. The maximal qubit value can be obtained from the following preparations and measurements: $\rho_x = (\openone + \vec{r_x}\cdot \vec{\sigma})/2$, $M_b^y = (\openone + b \vec{s_y}\cdot \vec{\sigma})/2$ with $\vec{s}_1=(\vec{r}_1+ \vec{r}_2)/\sqrt{2}$, $\vec{s}_2=(\vec{r}_1 - \vec{r}_2)
 /\sqrt{2}$, $\vec{r}_3=(-\vec{r}_1- \vec{r}_3)/\sqrt{2}$, and where $\vec{\sigma}=\{\sigma_x,\sigma_y,\sigma_z\}$ denotes the vector of Pauli matrices. Indeed, the correlation functions are then simply given by $E_{xy}=\vec{r}_x \cdot \vec{s}_y$.

An interesting feature of the witness $I_3$ is that it can also distinguish between classical and quantum resources of a given dimension; here, bits and qubits. If the inequality \eqref{322} (or one of its symmetries) is violated by a given probability distribution, then it follows that qubits, rather than classical bits, have been used. It is interesting to contrast this result with the Holevo bound \cite{Holevo}, which shows that one qubit cannot be used to send more than one bit of information. In our scenario, the state of the mediating particle somehow encodes the information about the value of the classical value $x$. However, here the use of quantum particles does provide an advantage.

Furthermore, we have strong numerical evidence that the following inequality (based on $I_3$) is never violated by qubits:
%We have also been able to find a 2-dimensional non-linear quantum witness based on $I_3$, which is of the form
\begin{eqnarray}\label{322Q}\nonumber
 J_3 \equiv | \arcsin{E_{11}}+\arcsin{E_{12}}+\arcsin{E_{21}} \\ -\arcsin{E_{22}} -\arcsin{E_{31}} | \leq \frac{3\pi}{2},
\end{eqnarray}
%We have numerical evidence that this inequality is never violated by qubits.
suggesting that $J_3$ may be used as a non-linear dimension witness. Moreover, the bound is tight, in the sense that there exist qubit preparations and measurements attaining it---for instance the states and measurements leading to $I_{3}=1+2\sqrt{2}$ given above.

\emph{2. Generalisation.} Next we generalise the witness $I_3$ presented above, in order to obtain classical and quantum dimension witnesses for any dimension. The form of $I_3$---see Eq. \eqref{322}---suggests the following natural generalisation for the case $N=m+1$:
\begin{eqnarray}\label{Nm}
 I_N \equiv \sum_{j=1}^{N-1}  E_{1j} + \sum_{i=2}^N  \sum_{j=1}^{N+1-i} \alpha_{ij} E_{ij} \\\nonumber
		\text{with} \,\, \alpha_{ij}=
		\begin{cases}
			+1& \text{if $i+j\leq N$},\\
            -1& \text{if $i+j= N+1$}.\\
		\end{cases}
\end{eqnarray}
It can be verified that for classical states of dimension $d\leq N$, the following relation holds:
\begin{equation}
I_N\leq L_d = \frac{N(N-3)}{2}+2d-1.
\end{equation}
Indeed for $d=N$ one obtains the algebraic bound $I_N=L_{d=N} = {N(N+1)}/2-1$. Using the methods of Ref. \cite{masanes2} we have checked that the inequality $I_N \leq L_{d=N-1}$ is a tight classical dimension witnesses ({\it i.e.} a facet of the polytope $\mathbb{P}_{N,m}^{d}$ with $m=d=N-1$) for $N\leq 5$.  Based on this evidence, we conjecture that it is a tight witness for all values of $N$.

Next we show that the inequality $I_N< L_{d=N}$ is a quantum dimension witness. More precisely, it is impossible to reach the algebraic bound of $I_N$ by performing measurements on quantum states of dimension $d=N-1$. Since $I_N$ is a linear expression of expectation values, it is sufficient to consider pure states, and one may write $E_{ij}= \bra{\psi_i} O_{j}\ket{\psi_i}$, where $O_j=M^j_{+1}-M^j_{-1}$ is the measured quantum observable. Clearly, in order to reach the algebraic maximum of $I_N$ we require $E_{ij}=\text{sign} [ \alpha_{ij} ]$ for $i+j\leq N+1$, and thus the states $\{\ket{\psi_i}\}$ must be eigenstates of the observables $\{O_j\}$ with eigenvalues $\{\text{sign} [ \alpha_{ij} ]\}$. From the structure of $I_N$, it can be seen that for any pair of preparations $\ket{\psi_s}$ and $\ket{\psi_t}$ with $1\leq s<t\leq N$, the observable $O_{N-t+1}$ must have eigenvalue $+1$ for $\ket{\psi_s}$ and eigenvalue $-1$ for $\ket{\psi_t}$. Thus all preparations must be mut
 ually orthogonal, since any pair of states $\ket{\psi_s}$ and $\ket{\psi_t}$ can be perfectly distinguished by measuring observable $O_{N-t+1}$. Since we must consider $N$ mutually orthogonal preparations, a Hilbert space of dimension (at least) $d=N$ is required to reach the algebraic maximum of $I_N$. It therefore follows that the inequality $I_N< L_{d=N}$ is a dimension witness for quantum systems of dimension $d=N-1$.

We believe, however, that better bounds can be obtained for the expression $I_N$. This is the case for $N=3$, as shown above, as well as for $N=4$ where we have been able to compute numerically the bounds for qubits and qutrits. These results are summarised in Table 1. Indeed, it would be desirable to find tight bounds for the witness $I_N$ for quantum states of any Hilbert space dimension $d<N$.
\begin{table}
	\begin{ruledtabular}
	\begin{tabular}{c||ccccc}
	&
	\textrm{$C_2$ (bit)}&
	\textrm{$Q_2$ (qubit) }&
	\textrm{$C_3$ (trit)}&
	\textrm{$Q_3$ (qutrit)}&
	\textrm{$C_4$ (quat)}\\
	\colrule
	$I_3$& 3 & $1+2\sqrt{2}$ & 5& 5&5\\
	$I_4$& 5 & 6 & 7& 7.9689 &9\\
	\end{tabular}
	\end{ruledtabular}
\caption{\label{tab:table1} Classical and quantum bounds for the dimension witnesses $I_3$ and $I_4$. Notably, these witnesses can distinguish classical and quantum systems of given dimensions.}
\end{table}

\textbf{\emph{Conclusion.}}---We have addressed the problem of testing the dimensionality of classical and quantum systems in a device-independent scenario. We have introduced the concept of `tight classical dimension witnesses' which allows one to put a lower bound on the dimensionality of classical states necessary to reproduce certain data. This naturally led us to generalise the concept of quantum dimension witnesses to arbitrary quantum systems. To illustrate these ideas, we have provided explicit examples of dimension witnesses. We have shown that these witnesses (i) are tight for small number of classical preparations, (ii) work both as classical and as quantum dimension witnesses, and (iii) allow one to distinguish classical and quantum states of given dimensions. Finally, we have introduced non-linear dimension witnesses, and have presented an example of such a witness for the simplest scenario. Furthermore, we believe that the simplicity of these techniques provides   sufficient appeal from the experimental viewpoint.

\emph{Acknowledgements.}---We acknowledge financial support from the European PERCENT ERC Starting Grant and Q-Essence project, the Spanish MEC FIS2007-60182 and Consolider-Ingenio QOIT projects, Generalitat de Catalunya and Caixa Manresa, and the UK EPSRC.

\end{document}